# Radiative cooling of colored paint based on $Fe^{3+}$ doped $Y_2Ce_2O_7$

Saichao Dang, Jingbo Xiang, Hongxin Yao, Fan Yang, Hong Ye*

Department of Thermal Science & Energy Engineering, University of Science and Technology of China, Hefei, Anhui 230027, P. R. China

*Corresponding to: H. Ye (hye@ustc.edu.cn)

**Abstract:** Materials with both low absorption of incoming solar radiation and high emittance in mid-infrared band can be applied for daytime radiative cooling. Current state-of-the-art materials for passive radiative cooling often utilize a combination of solar reflector and infrared emitter by different structures, or even by expensive nanofabricated photonic structures, which limits the applications in practice. In this study, possessing these two specified radiative properties, pure $Y_2Ce_2O_7$ is demonstrated with a performance of passive radiative cooling. With a bandgap at 375.7 nm, the prepared $Y_2Ce_2O_7$ shows a high solar reflectance of 91%, while with lattice strain and distortion of various bonds (e.g., Y-O, Ce-O), it also shows a high emittance of 0.96 in MIR band. More attracting, the aesthetics performance of $Y_2Ce_2O_7$ can be modified by doping $Fe^{3+}$ ions to change its color from ivory white to light yellow or red with high NIR reflection and MIR emission, indicating that the $Y_2Ce_{2-x}Fe_xO_7$ shows a better cooling performance than a common paint with a similar color. According to the field demonstration of cooling performance at noon time, the $Y_2Ce_2O_7$ and $Y_2Ce_{1.9}Fe_{0.1}O_7$ paints are 2.2 °C and 1.8 °C lower than the common white and umber paints, respectively, while at night, all paints are 2.3 °C lower than the ambient air. If applied on the envelop of a building, the simulation shows that the $Y_2Ce_2O_7$ and $Y_2Ce_{1.9}Fe_{0.1}O_7$ paints save 54.45% and 21.14% energy consumption compared with a common white and umber paints, respectively, in a hot season. The demonstrated $Y_2Ce_{2-x}Fe_xO_7$ holds potentials for energy-saving applications in hot climates.

**Keywords:** $Fe^{3+}$ doped $Y_2Ce_2O_7$, radiative cooling of colored paint, nanoparticle, Sol-Gel, energy-saving

## Introduction

Nowadays, survival and development of human being is threatened by global warming which could cause catastrophic climate change. The increasing energy consumption which brings vast greenhouse gases should take the main responsibility for this tough environmental problem because these gases will trap heat within the earth's atmosphere causing an increase in the temperature of the earth's surface over time [1-5]. With the fast development of urbanization in China, the energy consumption in building accounts for more than 23% of the total national energy consumption, and it is expected to increase further to 35% by 2020 [6-8]. As air conditioning occupies most of the energy consumption, without any external input energy, passive radiative cooling which uses the cold sky as a heat sink is a good technology for energy saving [9-12] in hot climates.

The nighttime radiative cooling has been extensively studied since 1975 [13-18]. As peak demand for cooling occurs during the daytime, it is more important to explore the possibility of passive daytime radiative cooling (DPRC). Efficient DPRC should have minimum absorption of incoming solar irradiation and maximum emission in the atmospheric transparency window (8~13 $\mu$m) for releasing heat to the cold sky [2, 4, 19]. A classic DPRC material consists of a bottom opaque metallic layer (e.g., Al or Ag) [9, 11, 12, 20, 21], which can reflect incident sunlight, and a rest top structure which shows a high emittance in the atmospheric transparency window and doesn't absorb sunlight as well. In these researches however, a definite nanofabrication process is required usually [10, 22]. A photonic 1D structure consisting of seven layers of $HfO_2$ and $SiO_2$ can achieve 4.9 °C lower than the ambient temperature with a cooling power of 40.1 W/m$^2$ when exposed to direct sunlight exceeding 850 W/m$^2$ on a rooftop [9]. Another demonstrated 2D array of symmetrically shaped conical metamaterial pillar structure consisting of seven periods of Al-Ge layers can possess a practical radiative cooling power of 116.6 W/m$^2$ [22]. By using polymers or nanoparticles, the radiative coolers can be economical [11, 20, 21, 23]. A dual-layer structure of a polytetrafluoroethylene film and a silver back reflector has been proposed [21]. A solar reflectance of 99% and a mid-infrared emittance of

90% are obtained, indicating a record-high day-time cooling power of 171.3W/m$^2$. Instead of using metal layer as a solar reflector, Bao [24] demonstrated double-layer coatings composed of densely packed TiO$_2$ particles which work as a solar reflective layer on top of densely packed SiO$_2$ or SiC nanoparticles which work as infrared emissive layer. These coatings can achieve 5 °C lower than the ambient temperature under direct solar radiation theoretically. Most of the DPRC researches are focused on the design of devices, such as materials with multilayer structures [9, 11, 21, 25, 26], photonic structures and metamaterials [22, 27, 28]. Their common characteristic is a combination of reflective layer and emissive layer and most of them have complicated structure and rely on expensive preparation technology.

To realize DPRC, the solar radiation should be reflected and the MIR emission should be enhanced. Instead of a combination of solar reflector and infrared emitter by different structures, a single material that can possess these two specified radiative properties will be a more attractive solution in practical applications. A new crystal Mg$_{11}$(HPO$_3$)$_8$(OH)$_6$ with both high reflectance (88.04%) in solar spectrum and high emittance (87.57%) in MIR is demonstrated for DPRC [5]. What's more, as the particles of Mg$_{11}$(HPO$_3$)$_8$(OH)$_6$ can be painted on the buildings in large scale, this crystal holds an attractive potential for building energy saving in hot climates. However, it shows only ivory white, indicating a poor aesthetic. High reflectance in visible will certainly make the pigments white color. However, white coatings make people's eyes uncomfortable for their reflected glare and also lack the aesthetics compared with darker colored coatings. Therefore, it is urgent to investigate radiative cooling of colored pigments for better practical application.

With high reflectance against the solar irradiation, several white inorganic pigments are demonstrated for energy-saving, such as $La_2Ce_2O_7$ [29], $La_2Mo_2O_9$ [30] and $Y_2Ce_2O_7$ [31]. with high reflection in NIR band, researchers have demonstrated some colored pigments, such as $Pr^{4+}$ and $Tb^{4+}$ doped $La_2Ce_2O_7$ [29], $Fe^{3+}$ doped $Y_2Ce_2O_7$ [31], $Fe^{3+}$ doped $YMnO_3$ [32]. Most of these work focus on the pigments' high reflection in solar spectrum. The radiative cooling performance in MIR band of such kind of pigments is neglected. As demonstrated in Ref [31], $Y_2Ce_2O_7$ shows a strong absorption in MIR band due to Ce-O and Y-O vibrations [33-38]. It is hoped that $Y_2Ce_2O_7$ can realize high emission in MIR band for radiative cooling in infrared band. Another neglected consideration of the above work is that the energy performance of colored paints which are made from those pigments in practical applications has not been demonstrated yet.

In this work, the $Fe^{3+}$ doped $Y_2Ce_2O_7$ pigments are synthesized by Sol-Gel method and characterized using several analytic techniques, such as SEM, XRD, UV-Vis-NIR spectrophotometer and FTIR. The chromatic properties of these pigments are studied. The mechanisms for their high reflection in solar band and high emittance in infrared band are investigated. Finally, the cooling performances of these pigments are discussed.

**Materials and Characterization**

As a typical preparation process, all pigments are synthesized by a Sol-Gel method. According to the formula of $Y_2Ce_{2-x}Fe_xO_{7+\delta}$ ($x$=0.05, 0.10, 0.20, 0.30), stoichiometric amounts of $Ce(NO_3)_3 \cdot 6H_2O$、$Y(NO_3)_3 \cdot 6H2O$ and $Fe(NO_3)_3 \cdot 9H_2O$ were dissolved into deionized water. As the solution was clear and with no residue, an amount of citrate acid (CA), which was used as the chelating agent to complex the cations, was added to the solution in the mole ratio 2:1 (CA: $Fe^{3+}$ = 2:1). After the CA was dissolved completely, a small amount of glucose (about 3 g in 100 ml) and acrylamide (in the mole ratio 4:1 with respect to the cations) were added to the solution. The particle size of the powder samples was controlled by acrylamide [39]. Subsequently, the PH value of the solution is adjusted to ~ 7 with aqueous ammonia. All the steps were done with magnetic stirring to ensure complete dissolution. The resultant solution was heated in water bath at 80 ◦C to initiate the polymerization reaction, and around 3 hours later, polyacrylamide gel was formed. Further the gel was dried at 120◦C in a temperature-constant oven for 12 h to obtain a dry gel precursor. The prepared gel precursor was grounded into powders and calcined at 900 ◦C for 4 h in a muffle furnace to produce $Y_2Ce_{2-x}Fe_xO_{7+\delta}$ powdered pigments [30, 31, 40].

The chemical composition formation was checked by energy dispersive spectrometer (EDS) analysis attached with SEM (Genimi SEM 500). The crystalline structure of the pigments was identified by X'Pert MPD XRD system using Cu-Ka radiation with continuous scanning mode at a rate of 12°/min ranges from 20° to 80°. Operating conditions of 40 kV and 15 mA were used to obtain the XRD pattern.

For the measurement of normal incident radiative properties of the pigments in solar spectrum and infrared band, a UV-visible-NIR spectrophotometers (SolidSpec-3700 DUV) and a Fourier transform infrared (FTIR) spectrometer (Bruker VERTEX 80) with an external integrating sphere were used, respectively. What's more, using KBr method, the transmission FTIR spectrum was recorded on Nicolet 8700.

The band gap energy (Eg) was obtained using the Kubulka-Munk reemission function [41] which is used as a measurement of absorption by powder. The function K-M is given by

$$F(R) = \frac{(1-R)^2}{2R} \tag{1}$$

where $R$ is the reflectance of sample. According to the relationship of band gap energy and absorption coefficient, there is:

$$[F(R)h\upsilon]^2 = B(h\upsilon - Eg) \tag{2}$$

Plotting $[F(R)h\upsilon]^2$ to $h\upsilon$ (Tauc plot) and linear extrapolating to F(R)=0, the intercept on the $h\upsilon$ corresponds to the band gap Eg.

# Results and discussion

## 3.1 The SEM result and XRD analysis

Figure 1a and 1b show the EDS spectra of $Y_2Ce_2O_7$ and $Y_2Ce_{1.7}Fe_{0.3}O_7$ pigments, respectively. The presence of all expected elements are identified. The inset SEM results show these coherent particles with a diameter of ~40nm. Figure 1c shows the XRD patterns of $Y_2Ce_{2-x}Fe_xO_{7+\delta}$ (x = 0.00, 0.05, 0.10, 0.15) powders. It is clear that all diffraction peaks belong to the standard cubic fluorite phase of $Y_2Ce_2O_7$, and doping of $Fe^{3+}$ does not change the structure of $Y_2Ce_2O_7$ phase.

## 3.2 The solar spectrum reflection, mechanism and chromatic properties

The measured solar reflection spectra of the $Y_2Ce_{2-x}Fe_xO_{7+\delta}$ ($x$ = 0.00, 0.05, 0.10, 0.15) pigments are shown in Figure 1d and the inset shows the corresponding Tauc plots. The white $Y_2Ce_2O_7$ pigments have high reflectance in solar spectrum. With the increase of Fe proportion, the absorption edge of $Y_2Ce_{2-x}Fe_xO_{7+\delta}$ compounds redshifts from 375.7 to 614.9 nm, and the corresponding band gap values decrease from 3.29 to 2.01 eV (see Table 1), and the doped samples thus appear light yellow and red (see Figure 2a). As the ionic radius of $Fe^{3+}$ (0.064 nm) is smaller than $Ce^{4+}$ (0.097 nm), the charge transfer transition between O(2p) valence and Fe(3d) conduction band of $Fe^{3+}$ will take a lower band gap than that transition between O(2p)- Ce(4f). The substitution of $Fe^{3+}$ for

$Ce^{4+}$ in $Y_2Ce_2O_7$ changes the absorption edge to a longer wavelength, i.e., a redshift, and as a result the reflectance of powders samples decrease with the amount of $Fe^{3+}$ increasing, meanwhile the color of samples changes from ivory white to light yellow and red.

The chromatic properties of the powders are shown in the CIE $L^* a^* b^*$ (1976) color space in Figure 2b. Each color in the CIE Lab color space has a unique location defined by its Cartesian coordinates with respect to the axes $L^*$, $a^*$ and $b^*$, where $L^*$ is the color lightness (0 for black, 100 for white), $a^*$ is the green (-)/red (+) axis and $b^*$ is the blue (-)/yellow (+) axis [42]. According to the visible reflectance of five compounds, their coordinate values are calculated and listed in Table 1. It is clear that with more $Fe^{3+}$ doped the value of $L^*$ decreases from 96.14 to 50.23, indicating the lightness of samples becomes dark as shown in Figure 2a. on the other hand, the increasing $a^*$ (from -0.15 to 14.34) and $b^*$ (from 6.90 to ~15) presents the redness and the yellowness of the pigments. As shown in Figure 2b, with the increasing of $x$ value, the color of $Y_2Ce_{2-x}Fe_xO_{7+\delta}$ moves from white to light yellow and red region. What's more, based on the visible reflectance of the pigments, their colors are calculated and plotted by Matlab software, as shown in Figure 2a. The calculations agree well with the colors of the pigments.

### 3.3 The infrared emittance and the mechanism

As shown in Figure 3a, the measured emittances of $Y_2Ce_{2-x}Fe_xO_{7+\delta}$ ($x$

= 0.00, 0.05, 0.10, 0.15) are very high and show a similar trend in 5~20 $\mu$m. When the wavelength is longer than 7.5 $\mu$m, these emittances are approximately 1, i.e., nearly a blackbody radiation. The doping of Fe doesn't change the high MIR emittance. Figure 3b shows the FTIR spectrum of $Y_2Ce_{2-x}Fe_xO_{7+\delta}$ ($x$ = 0.00, 0.05, 0.10, 0.15), which are obtained by KBr wafer method. The peak at 6.11 $\mu$m can be assigned to O-H bending vibration [38, 43]. Strong absorption bands are also observed between 7 and 13 $\mu$m. The band at 7.18 $\mu$m is attributed to Y-O vibration [33, 44] and the bands at 7.47 , 9.90 and 12.00 $\mu$m are attributed to the Ce-O vibration [34-36]. What's more, the Ce-O bond shows a series symmetry and asymmetry stretch vibrations[45] from 12.9 $\mu$m to 13.9 $\mu$m [36]. The compound of $Y_2Ce_{2-x}Fe_xO_{7+\delta}$ has a low symmetry, indicating a lot of infrared-active vibrational modes and higher infrared absorption[46]. The lattice strain and distortion due to the doping of Fe would further reduce the symmetry of the structure units, resulting in enhanced infrared radiation performance [47]. As the vibration of Fe-O bond may be overlapped with that of the Ce-O and Y-O bands, the influence of the doped Fe on the radiative properties in MIR is not obvious. In general, the $Y_2Ce_{2-x}Fe_xO_{7+\delta}$ ($x$ = 0.00, 0.05, 0.10, 0.15) show high emittance in MIR, indicating a great radiative cooling performance.

**3.4 The theoretical cooling performance**

To understand the cooling performance of five compounds, the radiative cooling power and equilibrium temperature are calculated according to an energy balance model as schematically shown in the inset of Figure 3c. Considering a structure of area $A$ at temperature $T$ with an ambient temperature of $T_{amb}$, the radiative cooling power ($P_{cool}$) can be obtained by

$$P_{cool}(T) = P_{rad}(T) - P_{atm}(T_{amb}) - P_{Sun} - P_{cond+conv} \qquad (3)$$

where $P_{sun}$ is the gained heat from the sun, which can be given by $P_{Sun} = A\int \alpha(\lambda, \theta_{Sun}) I_{AM1.5}(\lambda) d\lambda$. $I_{AM}1.5(\lambda)$ is the solar spectral irradiance at air mass 1.5 and $\alpha(\lambda, \theta_{Sun})$ is the absorbance in solar spectrum. Assuming the structure is facing the Sun at a fixed angle $\theta_{Sun}$, the term $P_{sun}$ does not have an angular integral, and the structure's absorbance is represented by its value at $\theta_{Sun}$. The power lost due to convection and conduction is

$$P_{cond+conv} = Ah_{com}(T_{atm} - T) \qquad (4)$$

with $h_{com}$ being the combined effective heat transfer coefficient. The structure will emit radiation and at same time receive radiation coming from the ambient in thermal infrared band. The radiated energy can be expressed as

$$P_{rad}(T) = A\int\int I_{BB}(T,\lambda)\varepsilon(\lambda,\theta)\cos\theta \, d\lambda \, d\Omega \qquad (5)$$

Here $\int d\Omega = 2\pi \int_0^{\pi/2} \sin\theta d\theta$ is the angular integral over a hemisphere. $\varepsilon(\lambda,\theta)$ is the spectral and angular emittance. $I_{BB}(T,\lambda) = \dfrac{2hc^2}{\lambda^5} \dfrac{1}{e^{hc/(\lambda k_B T)} - 1}$ is the spectral radiance of a blackbody at temperature T, where h is Planck's constant, $k_B$ is the Boltzmann constant, c is the speed of light. The absorbed power from the incident atmospheric thermal radiation is

$$P_{atm}(T_{amb}) = A \iint I_{BB}(T_{amb},\lambda) \alpha(\lambda,\theta) \varepsilon_{atm}(\lambda,\theta) \cos\theta d\lambda d\Omega \quad (6)$$

By using Kirchhoff's radiation law, structure's absorbance equals its emittance, i.e., $\alpha(\lambda,\theta) = \varepsilon(\lambda,\theta)$. The angle dependent emissivity of the atmosphere is given by

$$\varepsilon_{atm}(\lambda,\theta) = 1 - t(\lambda)^{1/\cos\theta} \quad (7)$$

where $t(\lambda)$ is the atmosphere transmittance in the normal direction and $\theta$ is the zenith angle [9]. In the present study, the high resolution atmosphere transmittance spectrum $t(\lambda)$ is taken from Ref [48] using a column water vapor of 1 mm and AM1.5. To estimate the cooling performance of the five compounds, the $P_{rad}(T)$ and $P_{atm}(T_{amb})$ are considered within 5~20 $\mu$m, assuming the emittance of the structures won't change as the temperature changes. The spectrum of solar radiation is taken from 0.28 to 2.5 $\mu$m. The averaged radiative properties can be defined as the integral of the properties. In MIR band, the averaged emittance can be obtained by

$$E_{\text{MIR}} = \frac{\int_{5}^{20} \varepsilon(\lambda) I_{\text{BB}}(\lambda) d\lambda}{\int_{5}^{20} I_{\text{BB}}(\lambda) d\lambda} \quad (8)$$

In solar spectrum, the averaged reflectance can be calculated by

$$R_{\text{solar}} = \frac{\int_{0.28}^{2.5} R(\lambda) I_{\text{AM1.5}}(\lambda) d\lambda}{\int_{0.28}^{2.5} I_{\text{AM1.5}}(\lambda) d\lambda} \quad (9)$$

Table 2 shows the averaged solar reflectance and MIR emittance of five compounds. Their $E_{\text{MIR}}$ are all very high and the $R_{\text{sol}}$ of the compound of $x=0$ is much higher than others', indicating a better cooling performance.

Assuming $T_{\text{amb}}=300$ K, the cooling power ($P_{\text{cool}}$) is calculated and plotted in Figure 3d as a function of structure temperature. The combined heat transfer coefficient is considered to be $h_{\text{com}} =0$, 6 and 12 Wm$^{-2}$K$^{-1}$. Obviously, the compound of $x=0$, i.e., white $Y_2Ce_2O_7$, shows a best cooling performance due to its high reflection in solar spectrum. At $T=300$ K, only this compound with a cooling power of 81 W/m² shows DPRC. The negative $P_{\text{cool}}$ means the structure is heated rather than cooled. At a temperature higher than 300 K, a positive cooling power can be realized by rest four colored compounds. If $T>300$ K, a higher $h_{\text{com}}$ causes a higher $P_{\text{cool}}$ due to the smaller $P_{\text{cond+conv}}$, while if <300 K, a higher $h_{\text{com}}$ is negative for radiative cooling, causing a lower $P_{\text{cool}}$. When temperature ($T$) goes up, $P_{\text{cool}}$ increases as well because a higher temperature means larger $P_{\text{rad}}$ and

$P_{cond+conv}$. At $T$=320 K, $h_{com}$=12 Wm$^{-2}$K$^{-1}$, five compounds of Y$_2$Ce$_{2-x}$Fe$_x$O$_{7+\delta}$ ($x$ = 0.00, 0.05, 0.10, 0.15) show cooling powers of 430, 185, 114, 72 and 13 W/m², respectively. To obtain colored coating for aesthetic, the price of energy consumption will be paid. By setting $P_{cool}$ =0, the equilibrium temperatures with different $h_{com}$ are shown in Figure 3c. The white Y$_2$Ce$_2$O$_7$ can be cooled 19 °C lower than the ambient ideally. With the increasing of $h_{com}$, the compounds will keep equilibrium temperatures close the ambient (300 K). With $h_{com}$=15 Wm$^{-2}$K$^{-1}$, the equilibrium temperatures of five compounds of Y$_2$Ce$_{2-x}$Fe$_x$O$_{7+\delta}$ ($x$ = 0.00, 0.05, 0.10, 0.20, 0.30) are 295.5, 308.1, 311.5, 313.6 and 316.3 K, respectively. With $x$ increasing from 0 to 0.3, the color of the compound becomes darker and the equilibrium temperature also increases 20.8 °C.

## 3.5 Cooling performance demonstration

Based on the above powders, several paints are prepared to demonstrate their cooling performance. Each paint consists of 85 wt% of the powder and 15 wt% of Polyvinylidene Fluoride (PVDF) as the binder. The mixture is dispersed in Nitrogen-methyl pyrrolidone (NMP, organic solvent) and disposed by ball-milling for 12 h. The mixture is painted on silica glass (2.5 cm×3.8 cm×1.0 mm). Smooth paint is obtained after drying in oven with 55 °C for 6 h to evaporate NMP. The prepared samples as shown in Figure 4a are exposed on rooftop to demonstrate their cooling performances. To reduce the influence of wind and humidity, the sample is covered by the polyethylene (PE) film and is surrounded by thermal insulator (styrofoam) as shown in Figure 4b. Al foil is applied to reduce the heat gain by the styrofoam. Temperature data at the surface of the samples are measured and recorded by real-time temperature recorder. For comparison, two common paints (i.e., white and umber) and an Aluminum are considered. Their radiative properties in solar band and MIR are shown in Table 2. Two colored paints show high MIR emittance while Al shows a low MIR emittance and a high solar reflection.

As shown in Figure 4c, the common white paint shows high temperature due to its low solar reflectance and Al foil also shows high temperature due to its low infrared emittance. The samples are not cooler than the ambient at daytime, which may because the small area of sample, the heat gain by the side walls of the sample and the high atmospheric humidity (70%) will weaken the radiative cooling performance. At noon time (11.30~14:30), the $Y_2Ce_2O_7$ paint is 2.2 °C lower than the common white paints. The cooling performance of the colored paints are shown in Figure 4d. The common umber paint shows high temperature due to its low solar reflectance. What's more, a paint with higher value of x also shows a higher temperature due to the larger absorption. With a similar color to the umber paint, the $Y_2Ce_{1.9}Fe_{0.1}O_{7+\delta}$ paint is 1.8°C lower than the common umber paint. As can be seen from Figure 4c and 4d, at night time, the paints are 2.3 °C lower than the ambient temperature due to their high infrared emittances.

## 3.6 Energy performance prediction

The paints are considered being applied on a building located in the city of Guangzhou (113° E, 23° N, China) to evaluate its energy performance. The four walls and the roof are coated by those paints, respectively. The room is 4.0 m in length, 2.8 m in height and 3.3 m in width. The south window is at the middle of the wall with a size of 1.5m×1.5m and it is considered to be a single-layer 6 mm-thick silica glass, of which the MIR emittance is 0.84 and the solar transmittance is 0.77. The indoor air temperature is maintained at 26 °C through space cooling as per the design standard for energy efficiency of China (JGJ 75–2003). The energy consumption for cooling the room can be calculated employing a building energy modeling program, BuildingEnergy. This program is an effective and reliable tool to analyze the building-related energy performance, and it has been validated in our previous work [7, 8]. According to this energy performance model, the energy consumption for cooling a room in Guangzhou in summer can be calculated. As shown in Figure 5, their energy performances in summer are compared.

Obviously, the white $Y_2Ce_2O_7$ shows a best energy performance. Compared with the common white paint, $Y_2Ce_2O_7$ saves 54.45% energy. The Al shows a slightly higher solar reflection than $Y_2Ce_2O_7$. However, duo to its low MIR emittance, the radiation heat transfer from Al to the ambient is suppressed and its energy consumption is 4 times that of $Y_2Ce_2O_7$. This fact emphasizes the importance of the regulation on MIR thermal radiation. With the increasing of Fe, the paints of $Y_2Ce_{2-x}Fe_xO_{7+\delta}$ ($x = 0.05, 0.10, 0.20, 0.30$) show a decreased solar reflection, causing more energy consumption for cooling. With a color coordinate of (69.42, 11.34, 16.64) in CIE $L^*a^*b^*$ space, the color of umber paint is close to the $Y_2Ce_{2-x}Fe_xO_{7+\delta}$ of $x=0.1$. The paint of $Y_2Ce_{1.9}Fe_{0.1}O_{7+\delta}$ consumes an energy of 85.00 kWh/m², which is 21.14% lower compared with the umber paint. Due to the purpose of the walls and roof, a better approach may be that the colored $Y_2Ce_{2-x}Fe_xO_{7+\delta}$ ($x = 0.05, 0.10, 0.20, 0.30$) paints are applied on the sides of a building for both esthetic and energy-saving and ivory white $Y_2Ce_2O_7$ is applied on the roof for radiative cooling.

## Conclusions

By Sol-Gel method, nanoparticles of $Y_2Ce_{2-x}Fe_xO_{7+\delta}$ ($x=0.00, 0.05$，0.10，0.20，0.30) are prepared for radiative cooling. Due to the charge transfer transition of O(2p)- Ce(4f) takes a band gap at 375.7 nm, the $Y_2Ce_2O_7$ compound shows a reflectance of 90.8% in solar spectrum. The doping of $Fe^{3+}$ ions reduces the solar reflection and changes the color of pigments from ivory white to light yellow and red, which is attributed to the charge transfer transition between O(2p) and Fe(3d). Various bonds (e.g., Y-O, Ce-O, Fe-O) are supposed to possess strong lattice strain and lattice distortion, which results high MIR emittance (>0.96). According to the field demonstration of cooling performance at noon time, the $Y_2Ce_2O_7$ paint is 2.2 °C lower than the common white paint and the $Y_2Ce_{1.9}Fe_{0.1}O_{7+\delta}$ paint is 1.8 °C lower than the common umber paint, while at night, all paints are 2.3 °C lower than the ambient temperature. If applied on the envelop of a building, the simulation shows that the $Y_2Ce_2O_7$ paint saves 54.45% energy consumption compared with a common white coating, and the $Y_2Ce_{1.9}Fe_{0.1}O_{7+\delta}$ paint saves 21.14% energy compared with an umber paint with a similar color in hot season. The demonstrated $Y_2Ce_{2-x}Fe_xO_{7+\delta}$ holds potentials for energy-saving applications in hot climates.

## Acknowledgements

This work was funded by the National Natural Science Foundation (No. 51576188). We thank Jinlan Peng and Xiuxia Wang at the USTC Center for Micro and Nanoscale Research and Fabrication for their help with characterization of SEM.

# Table Captions

Table 1 Color coordinates and band gap values of the $Y_2Ce_{2-x}Fe_xO_{7+\delta}$ (x = 0.00, 0.05, 0.10, 0.15) pigments.

Table 2 The averaged solar reflectance and MIR emittance of the $Y_2Ce_{2-x}Fe_xO_{7+\delta}$ (x = 0.00, 0.05, 0.10, 0.15) pigments and three coatings (white, umber and aluminum).

Table 1 Color coordinates and band gap values of the $Y_2Ce_{2-x}Fe_xO_{7+\delta}$ (x = 0.00, 0.05, 0.10, 0.15) pigments.

| Pigments | $L^*$ | $a^*$ | $b^*$ | Eg (eV) |
|---|---|---|---|---|
| x=0.00 | 96.14 | -0.15 | 6.90 | 3.29 |
| x=0.05 | 73.34 | 6.11 | 15.87 | 3.07 |
| x=0.10 | 64.23 | 9.19 | 14.37 | 2.96 |
| x=0.20 | 58.53 | 10.77 | 15.42 | 2.38 |
| x=0.30 | 50.23 | 14.34 | 16.39 | 2.01 |

Table 2 The averaged solar reflectance and MIR emittance of the $Y_2Ce_{2-x}Fe_xO_{7+\delta}$ (x = 0.00, 0.05, 0.10, 0.15) pigments and three coatings (white, umber and aluminum).

|  | $R_{sol}$ (%) | $E_{MIR}$ (%) |
|---|---|---|
| $x$=0.00 | 90.8 | 97.5 |
| $x$=0.05 | 63.2 | 97.1 |
| $x$=0.10 | 55.3 | 96.5 |
| $x$=0.20 | 50.5 | 97.4 |
| $x$=0.30 | 44.2 | 97.1 |
| White | 74.2 | 94.4 |
| Umber | 46.0 | 94.4 |
| Aluminum | 91.0 | 3.0 |

# Figure Captions

Figure 1 EDS spectra of Y2Ce2O7 (a) and Y2Ce1.7Fe0.3O7 (b) with the corresponding SEM result as inset. c) Powder X-ray diffraction patterns of Y2Ce2-xFexO7+$_\delta$ (x=0, 0.05, 0.1, 0.2, 0.3). d) Solar spectrum reflectance of Y2Ce2-xFexO7+$_\delta$ (x = 0.00, 0.05, 0.10, 0.15). Inset is the Tauc plots according to $K$–$M$ theory.

Figure 2 (a) The pigments of Y2Ce2-xFexO7+$_\delta$ (x = 0.00, 0.05, 0.10, 0.15) as well as the calculated color through the reflectance of the pigments. (b) Five pigments on CIE 1931 chromaticity space.

Figure 3 a) The MIR emittance of Y2Ce2-xFexO7+$_\delta$ (x = 0.00, 0.05, 0.10, 0.15). b) FTIR spectrum of Y2Ce2-xFexO7+$_\delta$ (x = 0.00, 0.05, 0.10, 0.15) obtained by KBr wafer method. (c) and (d) Theoretical cooling performance of the five pigments. c) Equilibrium temperature as a function of the combined heat transfer coefficient. d)

Figure 4 (a) and (b) Experimental demonstration of radiative cooling. a) Top view of the samples. b) Side view of one sample. Rooftop measurement of the temperatures of white paint, Al foil and paint of x=0 (c) and umber paint, x=0.05, x=0.1, x=0.2 and x=0.3 (d) against the ambient air temperature on a clear warm day in Hefei (117° E, 31° N, China) (03 February 2021).

Figure 5 Comparison of energy performance of eight coatings. The energy performance of an entire cooling period is considered, and the cooling period in Guangzhou starts from May 13th to October 17th of every year.

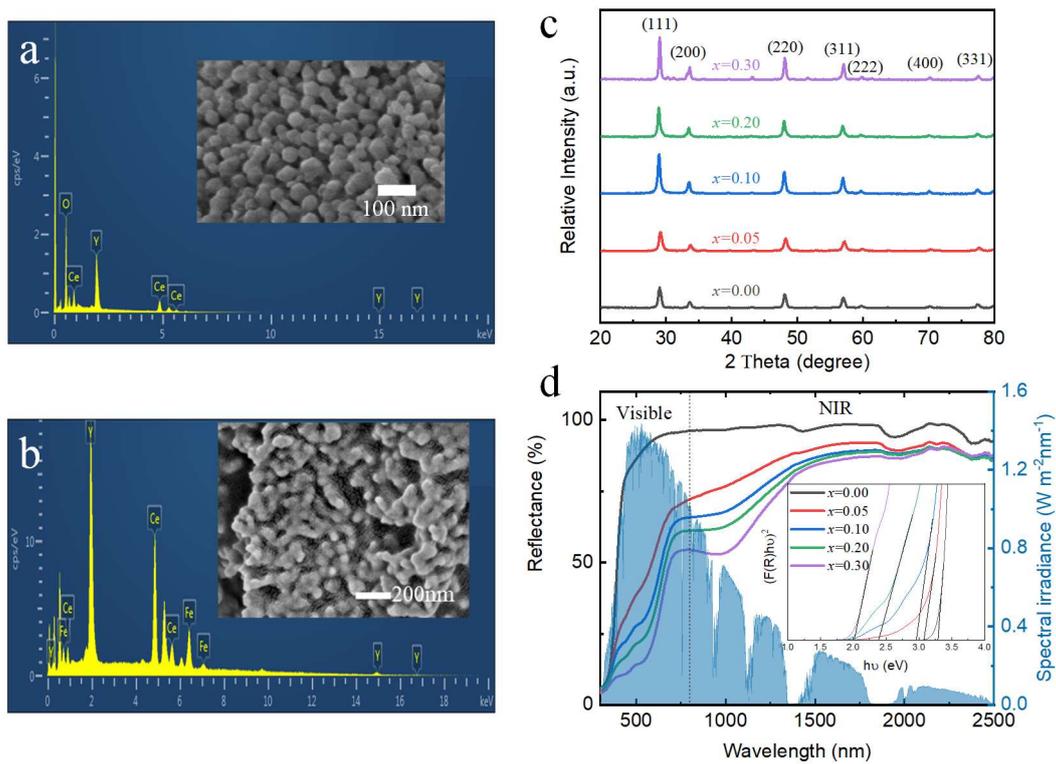

Figure 1 EDS spectra of $Y_2Ce_2O_7$ (a) and $Y_2Ce_{1.7}Fe_{0.3}O_7$ (b) with the corresponding SEM result as inset. c) Powder X-ray diffraction patterns of $Y_2Ce_{2-x}Fe_xO_{7+\delta}$ (x=0, 0.05, 0.1, 0.2, 0.3). d) Solar spectrum reflectance of $Y_2Ce_{2-x}Fe_xO_{7+\delta}$ (x = 0.00, 0.05, 0.10, 0.15). Inset is the Tauc plots according to K–M theory.

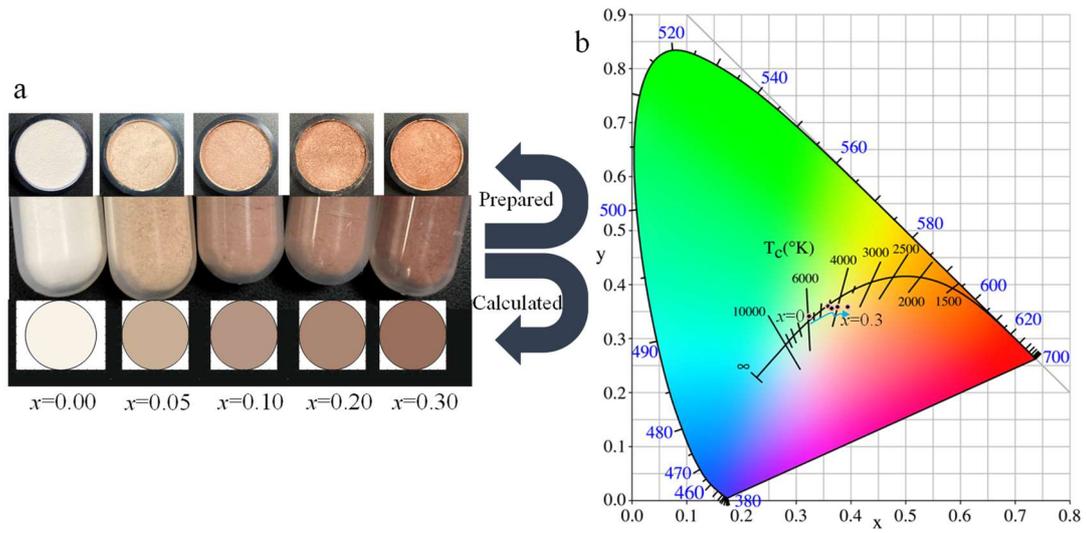

Figure 2 (a) The pigments of $Y_2Ce_{2-x}Fe_xO_{7+\delta}$ (x = 0.00, 0.05, 0.10, 0.15) as well as the calculated color through the reflectance of the pigments. (b) Five pigments on CIE 1931 chromaticity space.

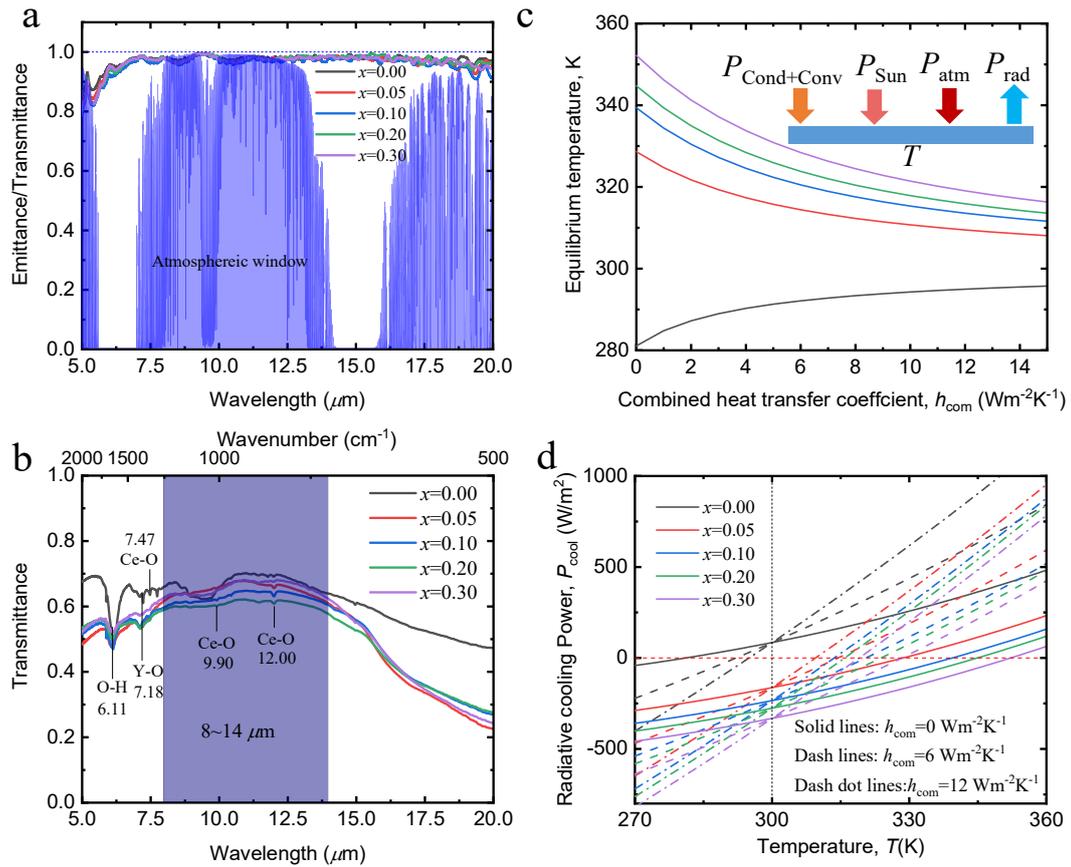

Figure 3 a) The MIR emittance of $Y_2Ce_{2-x}Fe_xO_{7+\delta}$ (x = 0.00, 0.05, 0.10, 0.15). b) FTIR spectrum of $Y_2Ce_{2-x}Fe_xO_{7+\delta}$ (x = 0.00, 0.05, 0.10, 0.15) obtained by KBr wafer method. (c) and (d) Theoretical cooling performance of the five pigments. c) Equilibrium temperature as a function of the combined heat transfer coefficient. d) Radiative cooling power as a function structure temperature.

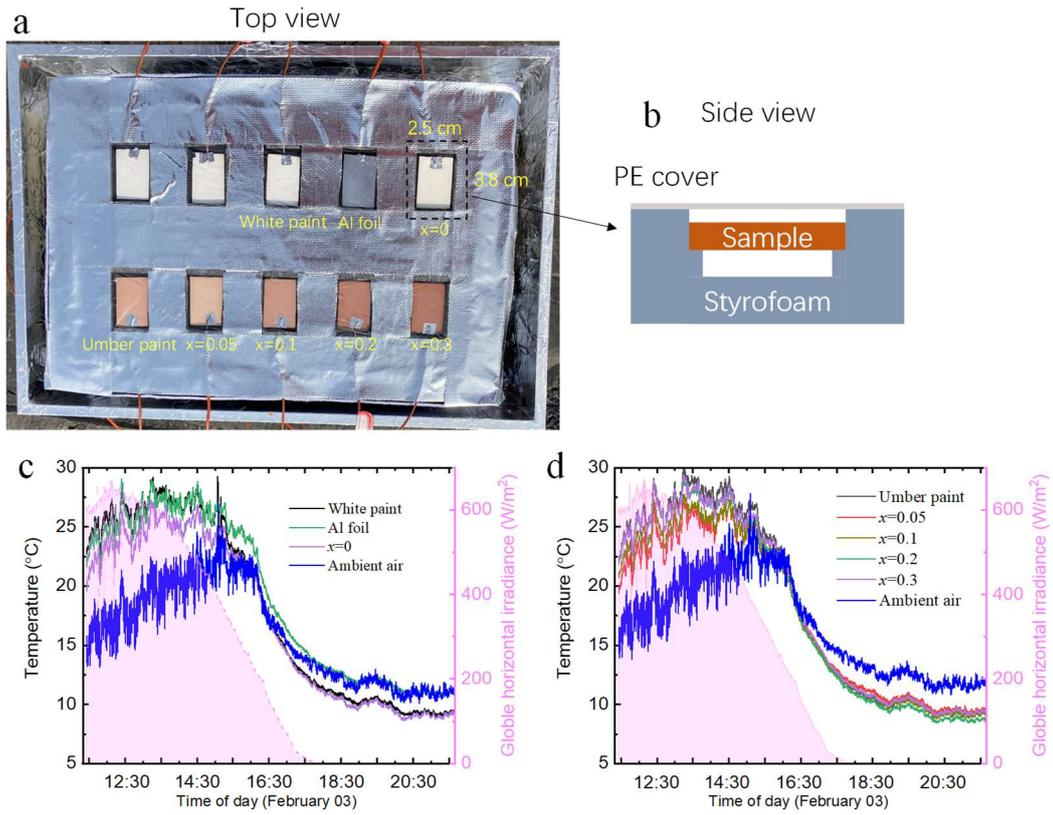

Figure 4 (a) and (b) Experimental demonstration of radiative cooling. a) Top view of the samples. b) Side view of one sample. Rooftop measurement of the temperatures of white paint, Al foil and paint of x=0 (c) and umber paint, x=0.05, x=0.1, x=0.2 and x=0.3 (d) against the ambient air temperature on a clear warm day in Hefei (117° E, 31° N, China) (03 February 2021).

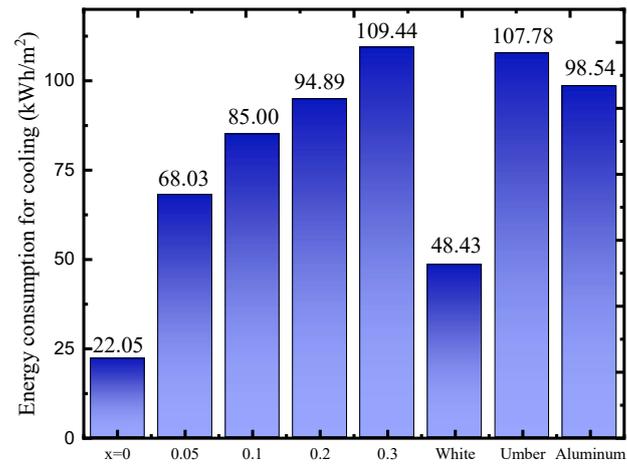

Figure 5 Comparison of energy performance of eight coatings. The energy performance of an entire cooling period is considered, and the cooling period in Guangzhou starts from May 13th to October 17th of every year.